\documentclass{elsart}
\usepackage{epsfig}
\usepackage{amssymb}

\begin{document}

\begin{frontmatter}

\title{Dynamics of the spin-boson model with a structured environment
\thanksref{birthday}}
\thanks[birthday]{This work is dedicated to Uli Weiss on the occasion of
his 60$^{\rm th}$ birthday.}

\author[delft,ntt]{{M.~Thorwart\corauthref{cor}}}
\corauth[cor]{Corresponding author}
\ead{thorwart@will.brl.ntt.co.jp}
\author[catania]{, E.~Paladino}
\author[delft]{, M.~Grifoni}

\address[delft]{Department of NanoScience, Delft University of
Technology, Lorentzweg 1, \\ 2628 CJ Delft, The Netherlands}
\address[ntt]{NTT Basic Research Laboratories, 3-1 Morinosato Wakamiya, \\ 
Atsugi-shi, Kanagawa 243-02, Japan}
\address[catania]{Dipartimento Metodologie Fisiche e Chimiche per
L'ingegneria, \\
Universit\'a di Catania, Viale Andrea Doria 6, 95125 Catania,
Italy \\
\& MATIS-Istituto Nazionale per la Fisica della Materia,
95123 Catania, Italy} 

\begin{abstract}
We investigate the dynamics of the spin-boson model when the spectral
density of the boson bath shows a resonance at a 
characteristic frequency $\Omega$  but behaves Ohmically at small 
frequencies. The time evolution of an initial state is determined by  
making use of the mapping onto a system composed of a 
quantum mechanical two-state system (TSS)
which is coupled to a harmonic oscillator (HO) with frequency $\Omega$. The
HO  
itself is coupled to an Ohmic environment. The dynamics is calculated 
by employing the numerically exact quasiadiabatic path-integral propagator 
technique. We find significant new properties compared to the 
Ohmic spin-boson model. By reducing the TSS-HO system in 
the dressed states picture to a three-level system for the special 
case at resonance, we  
calculate the dephasing rates for the TSS  analytically. 
Finally, we apply our model to experimentally realized 
superconducting flux qubits coupled to an underdamped 
dc-SQUID detector. 
\end{abstract}
\end{frontmatter}
%
% main text
\section{Introduction}
\label{sec.intro}
The spin-boson model is one of the central paradigms in theoretical physics
\cite{Weiss99,Leggett87}. The quantum mechanical two-state system (spin 1/2 or
in today's modern language qubit) interacting with a bosonic bath, 
is the simplest 
possible model to describe the effect of an environment on 
constructive and destructive quantum interference. It allows in 
a rather comprehensive way to investigate decoherence and damping 
on the quantum system imposed by a phenomenological environment. 
Hence, it has found 
numerous applications ranging from electron transfer to quantum 
information processing.

The characteristics of the harmonic bath in contact with the 
two-state system is captured by the spectral density $J(\omega)$ 
of the bath. 
A well studied special case is the spin-boson model with an Ohmic spectral 
density, where $J(\omega)\propto \omega$ (up to some cut-off 
frequency).  
This implies that the quantum system is damped 
 equally at all frequencies, which is the case for many physical systems. 
 A prominent example in the field of condensed matter physics are 
 single-charge effects in micro- and nanostructured systems in which 
 the electromagnetic environment can be modeled by an impedance of 
 pure Ohmic form. 
 The more general case of $J(\omega)$ having a power-law 
 distribution, i.e., $J(\omega)\propto \omega^s$ ($0<s<1$  sub-Ohmic and 
 $s>1$  super-Ohmic bath) also finds wide-spread applications, for instance
 when the TSS couples to an electromagnetic environment represented by a RC-transmission
 line. For these specifications, the time constants with which 
 a quantum superposition of the TSS decays can be calculated in analytic 
 form  by using real-time path integral techniques 
 \cite{Weiss99,Leggett87}. \\
 Another limit to be mentioned in this context 
  is the case of a single environmental mode at a specific 
 frequency $\Omega$, i.e., $J(\omega)\propto \delta(\omega-\Omega)$, which 
 is for instance realized by an electromagnetic environment with a 
 pure inductive impedance.  
 
 In this work, we consider the spin-boson model for the case of 
 the bath spectral density having a distinct resonance at a characteristic 
 frequency $\Omega$ of the order of the TSS level splitting, 
 and behaving Ohmically a low frequencies (peaked
 spectral density with Ohmic background).   This model 
 currently receives growing interest in the context of quantum computing 
 with condensed matter systems \cite{Makhlin01}, especially for superconducting 
 flux qubit devices \cite{vanderWaal00,ntt02,Chiorescu03,Semba03}. 
 In these devices, the state 
 of the TSS, i.e., the direction of the magnetic flux, is read out by a 
 dc-SQUID which couples inductively to the qubit superconducting loop. At
 bias currents well below the critical current $I_C$, the SQUID can be modeled 
 as an inductor $L_J$. An additional shunt capacitance $C_s$ creates an on-chip environment
 which improves the resolution of the dc-SQUID. The resulting impedance of
 this dc-SQUID environment can display a pronounced resonance at the characteristic
 frequency of the $LC$-circuit (see below for details). \\ 
%%%%%%%%%%%% Elisabetta
 As an alternative but equivalent point of view, a TSS coupled to a
 harmonic bath with a peaked spectral
 density represents an effective description of a TSS coupled to a damped
 harmonic oscillator of characteristic frequency $\Omega$.
 In this perspective, this model has recently been  used to describe solid 
 state implementations of a qubit coupled to a resonator.
 Various schemes of measurement, entanglement generation, quantum information
 transfer and interaction with a nonclassical state of the electromagnetic
 field have been proposed \cite{resonator,francescopino} and dephasing
 effects have been addressed \cite{francescopino}. 
 Moreover these devices, besides being promising candidates for realizing 
 qubits in a  possibly scalable quantum computer, are also very interesting 
 objects to display fundamental quantum physical properties, like macroscopic 
 quantum coherence \cite{Leggett85}, Rabi oscillations 
 \cite{Chiorescu03,Rabi37,Grifoni98}, Ramsey interference 
 \cite{Chiorescu03,Ramsey50}, etc. 

 The spin-boson model with a peaked spectral density of the 
 bath has been investigated
 theoretically in Refs.\ \cite{Thorwart00,Kleff03}. In Ref.\  
 \cite{Thorwart00}, it has been demonstrated that an external driving field 
 can be applied to slow down decoherence by moving the TSS out of 
 resonance with the HO due to dressing of the TSS by the driving 
 field. Recently, Kleff {\em et al.\/} \cite{Kleff03} have calculated for the 
 static case at zero temperature dephasing times by applying 
 the numerical flow equation renormalization method (see below for further
 remarks).
  
 To explore the more realistic finite-temperature 
 dynamics of the spin-boson model with a resonance peak 
 at frequency $\Omega$ in  the spectral density of the bath, 
 we make use of the already mentioned exact mapping to a TSS+HO model 
 \cite{Garg85}. By this, we obtain a model  
 consisting of the TSS  which couples to a harmonic oscillator 
 with characteristic frequency  $\Omega$, which itself couples 
 to a harmonic bath with {\em Ohmic\/} spectral density. Throughout this 
 work, we concentrate on the case of a weak coupling of the oscillator 
 to the Ohmic bath, but the TSS-HO coupling is kept arbitrary.  
 This choice of the parameter region is due to the experimentally given 
 situation of dc-SQUIDs which are typically underdamped. 
 This mapping allows to apply the 
 numerically exact method of the quasi-adiabatic 
 propagator path-integral (QUAPI) \cite{Makri92} for the TSS+HO being the
 the central quantum system. Ay additionally tracing out the HO degrees of
 freedom, we 
 determine the time-evolution of the reduced density matrix of the TSS. 
 We extract the dephasing and relaxation time constants for 
 the decay of a localized state of the TSS. In order to elaborate the 
 difference, we compare our numerical results to the well-known analytical expressions 
for the rates of a TSS coupled to a structured harmonic reservoir evaluated to lowest
order in the TSS-reservoir coupling  \cite{Grifoni97}. As it turns out,   
these 
 weak-coupling rates {\em deviate considerably\/} (up to a factor of 
 50) from the numerically exact decay rates for level splittings  around the 
 oscillator frequency. This shows that if the frequencies of the TSS and
 the HO are comparable,  the decay rates are determined by the 
{\em full\/} frequency spectrum of the peaked environment, and not only by its low
frequency weak Ohmic backround. Next, we concentrate on the specific case 
 of the TSS  being exactly in resonance with the oscillator and 
  establish a simple three-level description of the TSS-HO 
  system. Since we are interested in weak system-bath coupling, we 
  formulate the Redfield equations for this 
  three-level system (3LS) which can be solved analytically. The resulting 
  dephasing rates agree well with the numerically exact values of 
  QUAPI. Finally, we apply our general model to experimentally realized 
  superconducting flux qubit devices. 
\section{The model and the mapping to a TSS+HO Hamiltonian}  
To setup the model, we consider the Hamiltonian of a two-state system  
being coupled to a bath of (non-interacting) harmonic oscillators, i.e.,
\begin{eqnarray}
\tilde{H} & = & -\frac{\hbar\Delta_0}{2} \sigma_x - \frac{\hbar\varepsilon}{2} \sigma_z +
\frac{1}{2}\sigma_z \hbar \sum_k
\tilde{\lambda}_k (\tilde{b}_k^{\dagger} +\tilde{b}_k) + \sum_k
\hbar \tilde{\omega}_k \tilde{b}_k^{\dagger} \tilde{b}_k\, . \label{hamtot}
\end{eqnarray}
Here, $\tilde{b}_k$ and $\tilde{b}_k^{\dagger}$ are the annihilation and creation operators 
of the $k-$th bath mode with frequency $\tilde{\omega}_k$. 
The spectral density of the bath modes is assumed to have a Lorentzian peak
at the characteristic frequency $\Omega$, but behaves Ohmically at low 
frequencies with the dimensionless coupling strength $\alpha = \lim_{\omega \rightarrow
0} J(\omega)/2\omega$. The Lorentzian-shaped resonance has a width which 
we denote  as $\gamma=2\pi \kappa\Omega$. 
To be specific, we assume a spectral density of the form 
\begin{eqnarray}
J(\omega) & = & \sum_k \tilde{\lambda}_k^2 \delta(\omega-\tilde{\omega}_k)
=  \frac{2 \alpha \omega
\Omega^4}{(\Omega^2-\omega^2)^2+(2 \pi \kappa\omega \Omega)^2}
\mbox{\hspace{2ex}} \stackrel{\omega \rightarrow 0}{\longrightarrow} 2
\alpha \omega \, .
\label{jeff}
\end{eqnarray}
For our purposes, it is convenient to aquire a different viewpoint as it
will become clear below.
Following Ref.\ \cite{Garg85,Thorwart00}, there exists an exact one-to-one mapping of
this Hamiltonian onto that one of a TSS coupled to a single HO 
 mode with frequency $\Omega$ which itself interacts with a set of 
(non-interacting) harmonic 
oscillators having an {\em Ohmic\/} spectral density with the dimensionless damping
strength $\kappa$. Thereby,  the interaction strength between the TSS and the HO is 
given by $g$. The corresponding Hamiltonian reads
\begin{eqnarray}
H & = & -\frac{\hbar \Delta_0}{2} \sigma_x - \frac{\hbar\varepsilon}{2} \sigma_z +
\hbar g \sigma_z (B^{\dagger} + B) + \hbar \Omega B^{\dagger} B \nonumber \\
& & + (B^{\dagger} + B) \sum_k
\hbar \nu_k (b_k^{\dagger} + b_k) + \sum_k
\hbar \omega_k b_k^{\dagger} b_k \, . \label{hamtlsosc}
\end{eqnarray}
Here, $B$ and $B^{\dagger}$ are the annihilation and creation operators 
of the localized HO mode while the $b_k$ and the $b_k^{\dagger}$ are
the corresponding bath mode operators. The spectral density of the 
continuous bath modes now becomes Ohmic, i.e., 
\begin{equation}
J_{\rm Ohm}(\omega) = \sum_k \nu_k^2 \delta(\omega-\omega_k) 
= \kappa\omega e^{-\omega / \omega_c} \, , 
\label{johm}
\end{equation}
where we have introduced the usual high-frequency cut-off at $\omega_c$ 
(We note that throughout this work, we use the value 
$\omega_c=10 \Delta_0$). 
The relation between $g$ and $\alpha$ follows as 
\begin{equation}
g=\Omega \sqrt{\frac{\alpha}{8 \kappa}} 
\hspace{3ex} \Leftrightarrow \hspace{3ex}  \alpha = 8 \kappa \frac{g^2}{\Omega^2} \, . 
\end{equation}
Since for our cases of interest, the damping of the HO is small ($\kappa
\ll 1$), we will denote the TSS+HO-system as the central quantum system 
 which is itself weakly damped. However, the coupling between 
the TSS and the HO can still become  large. In the following, we will
evaluate the reduced density operator 
$\rho_{TSS+HO}(t)=tr_{\rm bath} e^{i H t/\hbar} W(0) e^{-i H t/\hbar}$
 for the TSS+HO-system, where the bath degrees of freedom have been traced
 out.  $W(0)$ denotes the full density 
 operator at initial time $t=0$. Throughout this work, 
 we assume a factorizing initial preparation where all three parts of the 
 total system are decoupled and the coupling is switched on instantaneously 
 at $t=0^+$. In particular, we choose the bath being at thermal equilibrium
 at inverse temperature $\beta=1/ k_B T$ as well as  the localized mode
 being thermally distributed. This implies
\begin{equation}
W(0) = \rho_{TSS}(0) \otimes \frac{e^{-\beta H_{HO}}}{Z_{HO}} \otimes
\frac{e^{-\beta H_{bath}}}{Z_{bath}}  \;, 
\end{equation}
where $Z$ denotes the partition function of the corresponding subsystem. 
 The further reduction to the reduced
density operator $\rho(t)$ of the TSS alone is easily done by tracing out the HO
degrees of freedom, i.e., $\rho(t)=tr_{HO} \, \rho_{TSS+HO}(t)$. 
%%%%%%%%%%%%%%%%%%%%%%%%%%%%%%%%%%%%%%%%%%%%%%%%%%%%%%%%%%%%%%%%%%%%%%%%%%%%%%%%%%%%%%%%%%
%%%%%%%%%%%%%%%%%%%%%%%%%%%%%%%%%%%%%%%%%%%%%%%%%%%%%%%%%%%%%%%%%%%%%%%%%%%%%%%%%%%%%%%%%%
\section{Observables}
Having calculated the reduced density matrix  $\rho(t)$ of the TSS, we can 
directly calculate the observables of interest. In this context, we are 
interested in the dephasing and relaxation rates of the TSS  dynamics. 
Hence, we prepare the TSS  in a symmetric superposition  
of its energy eigenstates. 
The relevant dynamical quantity then is the symmetrized correlation function 
\begin{equation}
S_z(t) = \frac{1}{2} \langle \sigma_z(t) \sigma_z(0)+\sigma_z(0) 
\sigma_z(t)\rangle - \langle \sigma_z \rangle^2_{\infty,+}\, , 
\end{equation}
where $\sigma_z(t) = e^{i H t/ \hbar}\sigma_z e^{-i H t/ \hbar}$ and where
the equilibrium population $\langle \sigma_z \rangle^2_{\infty,+}$ 
(the subscript $+$ indicates
that the dynamics has been calculated with positive bias $\varepsilon >
0$) 
has been subtracted. By assuming a factorizing
initial preparation, the correlation function can be 
expressed in terms of the population difference 
$P_+(t) \equiv \langle \sigma_z \rangle_t=P_{s}(t) + P_{a} (t)$. It follows
\begin{equation}
S_z(t) = P_s(t) + P_{\infty, +} [ P_a(t) - P_{\infty, +}] \, ,
\label{szt2}
\end{equation}
where $P_{s/a} (t)$ are the symmetric/antisymmetric parts of $P_+(t)$ with
respect to the sign of the bias and $P_{\infty, +} \equiv 
\langle \sigma_z \rangle^2_{\infty,+}$. This quantity vanishes at long times which
ensures that the Fourier transform exists. By calculating the population
$P_-(t)$ with the sign of the bias inverted ($\varepsilon < 0$), 
we obtain $P_{s/a}(t)$ according to 
\begin{eqnarray}
P_s(t) & = &  [P_{+}(t) + P_{-} (t)]/2 \, , \nonumber \\
P_a(t) & = &  [P_{+}(t) - P_{-} (t)]/2 \, . 
\label{pas}
\end{eqnarray}
For the unbiased case $\varepsilon=0$, we have $S_z(t) = P_s(t)\equiv
P(t)$. Moreover, the initial preparation $\rho_{TSS}(0)$ of the 
TSS in an equally weighted superposition of ground and excited state implies 
$P_+(0)=P_-(0)=1$. 
Since $S_z(t)$ vanishes at $t \rightarrow \infty$, it readily can be
Fourier transformed, i.e., 
\begin{equation}
S_z(\omega) = 2 \int_0^\infty dt \cos \omega t \, S_z(t)  \, . 
\label{szom}
\end{equation}
To extract the decay rates $\Gamma_i$, 
the frequencies $\omega_i$ and the spectral weights $a_i$ associated to
 $S_z(\omega)$, 
we approximate  $S_z(t)$  as a sum of exponentially 
decaying sinusoids, i.e., 
\begin{equation}
S_z(t) = \sum_i a_i e^{-\Gamma_i t} \cos {\omega_i t} \, ,
\end{equation}
where $\sum_i a_i =1$. 
It has the Fourier transform 
\begin{equation}
S_z(\omega) = \sum_i a_i \left[\frac{\Gamma_i}{\Gamma_i^2 + (\omega_i+\omega)^2} 
+\frac{\Gamma_i}{\Gamma_i^2 + (\omega_i-\omega)^2} \right]
\;. \label{fit}
\end{equation}
By a standard fitting procedure, we fit the exact numerical results to this
function to extract the decay rates $\Gamma_i$ (HWHM of the Lorentzian peak), 
the frequencies
$\omega_i$ (position of the peak) and the spectral weights $a_i$. 
%
%%%%%%%%%%%%%%%%%%%%%%%%%%%%%%%%%%%%%%%%%%%%%%%%%%%%%%%%%%%%%%%%%%%%%%%%%%%%%%%%%%%%%
%%%%%%%%%%%%%%%%%%%%%%%%%%%%%%%%%%%%%%%%%%%%%%%%%%%%%%%%%%%%%%%%%%%%%%%%%%%%%%%%%%%%%
\section{The numerical ab-initio technique QUAPI}
In order to calculate the reduced density operator, we use 
the iterative tensor multiplication scheme derived 
for the so-called quasiadiabatic propagator path integral (QUAPI). 
This 
numerical algorithm was developed by Makri and Makarov \cite{Makri92} within 
the context of chemical physics. Since its first applications it has been 
successfully tested and 
adopted to various problems of open quantum systems, with and without 
external driving \cite{Makri92,ThorwartQUAPI,ThorwartQUAPI2,Pechukas99}. 
The details of this technique have been discussed to a great extent in the literature 
 \cite{Makri92,ThorwartQUAPI,ThorwartQUAPI2,Pechukas99} and 
 we only mention briefly how to adopt it to our specific 
 problem. 
 
\begin{figure}[t]
\begin{center}
\epsfig{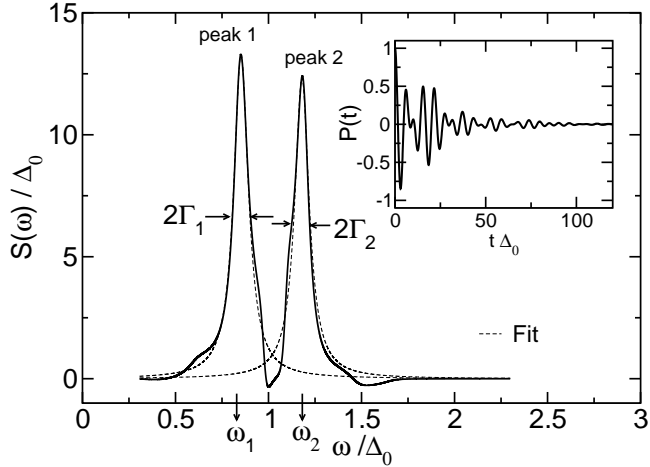}
\caption{Example of the dynamics for the symmetric case $\varepsilon=0$, 
where the oscillator frequency is in resonance with the TSS frequency,
i.e., $\Omega=\Delta_0$. Parameters are: $g=0.18 \Delta_0$, $\kappa=0.014 (\rightarrow
\alpha = 0.004), k_{B}T=0.1 \hbar \Delta_0$. QUAPI parameters are
$M=12, K=1, \Delta t=0.06 /\Delta_0$. 
\label{fig.ex}}
\end{center}
\end{figure}

We use QUAPI to calculate the reduced density matrix $\rho(t)$ 
and iterate until the stationary 
state of $P_{\pm}(t)$ has been reached. For the biased case, we run the iteration with 
both positive and negative bias 
to calculate $P_{\pm}(t)$ ($\varepsilon >0$ and $\varepsilon <0$). 
Having transformed to $P_{s/a}(t)$, see Eq.\  (\ref{pas}), 
we determine $S_z(t)$  in Eq.\  (\ref{szt2})
and Fourier transform the result to  
determine $S_z(\omega)$ according to Eqs.\  (\ref{szom},). 
Then, a fit 
to Eq.\  (\ref{fit}) determines the dephasing and relaxation rates 
$\Gamma_i$ (HWHM of the Lorentzian peak), 
the frequencies
$\omega_i$ (position of the peak) and the spectral weights $a_i$. 
A typical result for the symmetric TSS ($\varepsilon=0$) being in resonance
with the HO ($\Delta_0=\Omega$) is shown 
in Fig.\  \ref{fig.ex}, see also Section \ref{3ls} for a more detailed
discussion of this specific case. The inset of Fig.\  \ref{fig.ex} depicts 
the time evolution of $P(t)$. The Fourier transform displays two 
characteristic peaks at $\omega=\omega_1$ and $\omega=\omega_2$. 
The width of these peaks can be read of the fitted Lorentzians, see dashed
lines. 

\begin{figure}[t]
\begin{center}
\epsfig{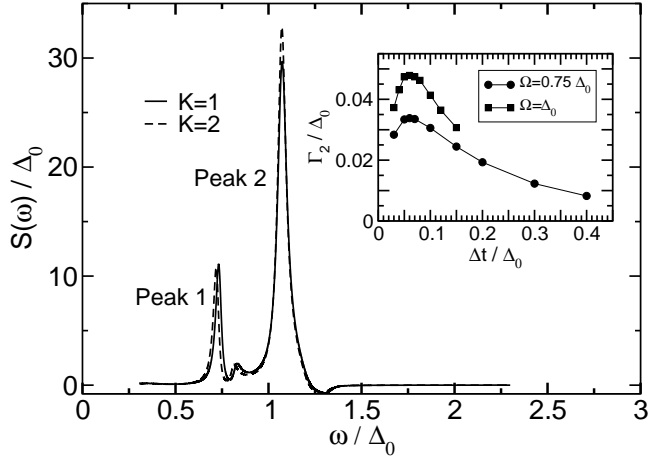}
\caption{Main: Fourier transform $S_z(\omega)$ for two different memory 
lengths $K=1$ and $K=2$, see text. Parameters: $\varepsilon=0, g=0.18 \Delta_0$,
 $\Omega= 0.75 \Delta_0, \kappa=0.014  (\rightarrow
\alpha = 0.006), k_{B}T=0.1 \hbar
\Delta_0$. Remaining QUAPI parameters are
$M=12, \Delta t=0.06 /\Delta_0$. Inset: Dephasing rate $\Gamma_2$ for
different QUAPI-time steps $\Delta t$ for two different values of 
$\Omega$. The position of the 
maxima determine the choice of $\Delta t$ for QUAPI, see text. 
\label{fig.vark}}
\end{center}
\end{figure}

 QUAPI uses three free parameters which have to be adjusted 
 according to the particular situation. Typical physically 
 reasonable situations allow for a tractable choice of parameters. \\
  (i) The first parameter is the 
 dimension $M$ of the Hilbert space of the quantum system which is 
 in our case the composed TSS+HO-system. With the dimension of the TSS 
 being fixed, we truncate the in principal infinite dimensional oscillator 
 basis to  the six lowest 
 energy eigenstates leading to $M=12$. We have checked convergence also 
 with $M=16$ and $M=20$ and found the results unchanged.   \\
 (ii) The second parameter affects the length of the memory which 
 is induced by the environmental correlations. This memory time has the
 length of $K$ times the time step of the iteration. 
 Here, we deal with weak Ohmic
damping ($\kappa\ll 1$) of the HO and temperature in the experimental devices 
is typically larger than damping. 
(This in principle
would even allow for a Markovian approximation.) In our case, this permits
to choose $K$ small. We fix $K=1$. We have also ensured convergence 
by choosing $K=2$, but the results changed only by less than 10 \%. 
This is depicted in Fig.\  \ref{fig.vark}. The results for peak 1 are for
$K=1: \Gamma_1=0.0177, \omega_1=0.730, a_1=0.197$ (in  units of $\Delta_0$) 
while for $K=2: \Gamma_1=0.0166, \omega_1=0.718, a_1=0.181$. 
For peak 2, we find the values $K=1: \Gamma_2=0.0310, \omega_2=1.074,
a_2=0.913$, while for $K=2: \Gamma_2=0.0282, \omega_2=1.073, a_2=0.917$.
The sum rule $a_1+a_2=1$ is fulfilled with an accuracy of
approximately $10 {\%}
$. \\
At this point, it becomes clear why it is essential for QUAPI to use the
mapping to the TSS+HO model.  For the dynamics of
the spin-boson model with the peaked spectral density in Eq.\  
(\ref{jeff}), the autocorrelation function $L(t)=\langle \xi(t)
\xi(0) \rangle, t> 0$, of the fluctuating  force $\xi(t)$  of the bath \cite{Weiss99}
decays slowly with time in an oscillatory way. This can be readily
understood since $L(t)$ is qualitatively the Fourier transform of
$J(\omega)$. In other words, the peaked environment possesses a longer 
correlation time compared to the Ohmic bath. This is also the reason why
a standard weak-coupling approach fails to determine the dephasing rates 
(see  Sec.\ \ref{Sec.weakcoupl} for a detailed comparison between both
approaches). Since the long correlation time of the peaked bath 
 would require a long memory time 
$K\Delta t$ with $K\gg 1$, the direct treatment of the TSS coupled to the
peaked bath with QUAPI is still beyond the today's computational 
resources. Only the corresponding Ohmic autocorrelation function decays
fast enough with time to be truncated after a relatively short memory time
(at any finite temperature $T$). It is computationally more favorable to
iterate the HO dynamics as a part of the quantum system and not as a part
of the bath. \\
We mention that for the comparison with the three-level approximation (see
Section \ref{3ls} below), which is expected to be good at low temperatures, we used the
parameters $M=6$ and $K=3$, since in this case the bath-induced
correlations have a weaker decay  and hence require a longer memory time. 
Also, the lower temperature allows to restrict to less basis states.  \\
(iii) The third QUAPI-parameter is the time-step $\Delta t$. 
It is fixed according to the principle of minimal
sensitivity \cite{ThorwartQUAPI2} of the result on the variation 
of  $\Delta t$.  The results  which permits to decide on
$\Delta t$ are shown in the inset of Fig.~(\ref{fig.vark}). The position 
of the maxima determine the choice of $\Delta t$. In our case, we fix 
$\Delta t=0.06 / \Delta_0$. As one can observe from  
Fig.~(\ref{fig.vark}), the position of the maximum is not sensitive on the 
change of the parameter $\Omega$. 
\section{Varying the oscillator frequency $\Omega$}
In this section, we investigate the effect of varying the frequency
$\Omega$ of the resonance of the spectral density $J(\omega)$ in Eq.\
(\ref{jeff}),  or, in other words, the 
frequency of the harmonic oscillator. For simplicity, we restrict 
here to the symmetric TSS $\varepsilon=0$. For the biased TSS, the 
qualitative behavior is similar if one replaces $\Delta_0$ by the effective
level splitting 
$\Delta_b= \sqrt{\Delta_{0}^2 + \varepsilon^2}$ for the biased case.

\subsection{Exact resonance $\Omega = \Delta_0$ \label{sec.res}}
The dynamics of the TSS being in resonance with the HO is depicted in Fig.\
   \ref{fig.ex} for a typical set of parameters (see caption). 
   One observes a decay of $P(t)$ with two frequencies and
   two damping constants. This can be readily understood by considering 
   the undamped TSS+HO-system, i.e., $\kappa=0$. To diagonalize the
   corresponding Hamiltonian, we can use the rotating-wave approximation
   (RWA), which is appropriate at resonance. Treating the interaction $g$ in
   first order perturbation theory, we obtain the spectrum as a ladder of 
   energy-levels. The groundstate $|g,0\rangle$ ($g$ denotes the
   groundstate of the TSS while $0$ denotes the index of the HO mode) is
   separated by $n \hbar\Omega$ from the higher lying n-th pair of excited 
   states which are almost
   degenerate, but split by $\hbar\overline{\Delta}_n=2\hbar g \sqrt{n}$. At rather low 
   temperature, only the three lowest energy eigenstates, i.e., the 
   ground state  $|g,0\rangle$ and the first ($n=1$) pair of excited states are
   relevant. This yields to two peaks in the spectrum at
   $\omega_1=\Delta_0-g$ and at $\omega_2=\Delta_0+g$ with the distance
   $\Delta_1=2g$
   which appears in Fig.\  \ref{fig.ex}. The two peaks have almost the same
   height, i.e., the same spectral weight $a_1 \approx a_2$.

\subsection{Negative detuning $\Omega  < \Delta_0$ \label{subsec.negdet}}
\begin{figure}[t]
\begin{center}
\epsfig{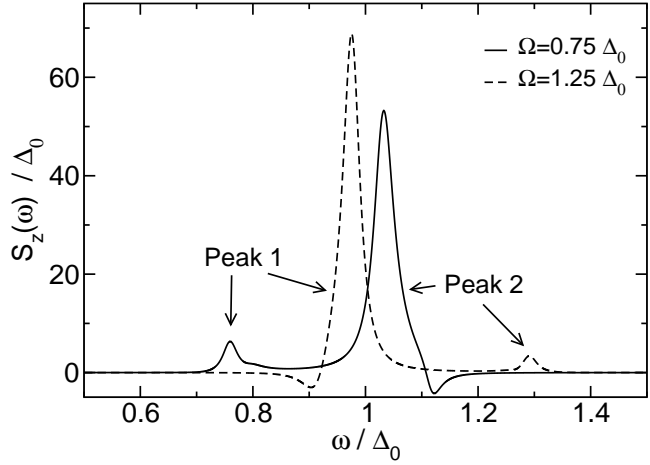}
\caption{$S_z(\omega)$ for two values of the oscillator frequency 
$\Omega$. Parameters are: $\varepsilon=0, g=0.07 \Delta_0$, $\kappa=0.014, k_{B}T=0.1 \hbar
\Delta_0$. 
\label{fig.varom1}}
\end{center}
\end{figure}

If the HO frequency is detuned from the TSS level splitting such that 
$\Omega < \Delta_0$, the dynamics of $P(t)$ also contains two frequencies
but with different spectral weights. This is shown in Fig.\  
\ref{fig.varom1} for $\Omega=0.75 \Delta_0$. 
Also, a perturbative analysis of second order in $g$ 
(the first order vanishes exactly), readily explains the features: 
again restricting to the three lowest energy eigenstates, we find that the 
first excited state $|g,1\rangle$ is separated from the ground state
 $|g,0\rangle$ by $\omega_1=\Omega-\frac{2g^2\Delta_0}{\Delta_0^2-\Omega^2} \approx 
 \Omega-\frac{g^2}{\Delta_0-\Omega}$, where in the last relation the RWA
 has been used. The almost degenerate next higher lying excited state 
 $|e,0\rangle$ is separated from the ground state by 
 $\omega_2=\Delta_0+\frac{2g^2\Delta_0}{\Delta_0^2-\Omega^2} \approx 
 \Delta_0+\frac{g^2}{\Delta_0-\Omega}$. These two frequencies show up in the 
 spectrum, see  Fig.\ \ref{fig.varom1}. The perturbatively calculated
 values of the peak positions are for these parameters $\omega_1=0.73
 \Delta_0$ and $\omega_1=1.02 \Delta_0$. 
 The spectral weights of the peaks are different, i.e., $a_1 \ll a_2$. 
\subsection{Positive  detuning $\Omega  > \Delta_0$}
For the case of positive detuning $\Omega  > \Delta_0$, also two
frequencies appear, see Fig.\  \ref{fig.varom1}. 
A perturbative treatment as in Subsection 
\ref{subsec.negdet} gives the frequency of the peaks at 
$\omega_1=\Delta_0+\frac{2g^2\Delta_0}{\Delta_0^2-\Omega^2} \approx 
 \Delta_0+\frac{g^2}{\Delta_0-\Omega}$ and 
 $\omega_2=\Omega-\frac{2g^2\Delta_0}{\Delta_0^2-\Omega^2} \approx 
 \Omega-\frac{g^2}{\Delta_0-\Omega}$, where in the last relations the RWA
 has been used again. For the parameters used in Fig.\ \ref{fig.varom1}, 
 the results are $\omega_1=0.98$ and $\omega_2=1.27$ which agree 
 well with the positions of the peaks in the spectrum. 
  The spectral 
 weights are in this case inverted, i.e.,  $a_1 \gg a_2$. 
\subsection{Dephasing rates}
\begin{figure}[t]
\begin{center}
%\hfill
\epsfig{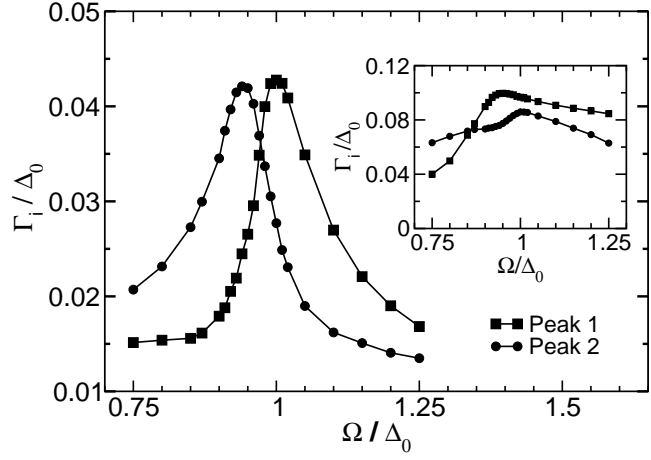}
\caption{Main: Dephasing  rates corresponding to peak 1 and peak 2 in the
Figs.\  \ref{fig.ex} and \ref{fig.varom1} as a function of the HO frequency
$\Omega$. The parameters are: $\varepsilon=0, g=0.07 \Delta_0$,
$\kappa=0.014, k_{B}T=0.1 \hbar
\Delta_0$. Inset: Same for stronger damping $\kappa=0.02$ with $\alpha=0.01=$const.  
(like in Ref.\ \cite{Kleff03}). This implies that with varying $\Omega$ also $g$ is
changed. \vspace{3mm} \label{fig.varom2}
}
\end{center}
\end{figure}

In order to extract the dephasing rates $\Gamma_i$, we fit two Lorentzians
to the peaks in the spectrum shown in Figs.\  \ref{fig.ex} and 
\ref{fig.varom1}. The dephasing rates follow according to Eq.\  
(\ref{fit}). The dependence of the 
dephasing on the frequency of the HO is shown in Fig.\  \ref{fig.varom2}. 
The main figure displays the result if the interaction
strength $g$ between TSS and HO is kept fix, while its inset 
depicts the result for $\alpha$ kept fix, i.e., $g$ also varies with 
varying $\Omega$. Each figure displays the dephasing rates of both peaks. 
The dominating peak with the dominant spectral weight is peak 2 for 
negative detuning $\Omega< \Delta_0$, while it is peak 1 for positive 
detuning $\Omega > \Delta_0$. Both peaks show a clear maximum at which the
dephasing is largest. This happens naturally in the vicinity of the 
resonance $\Omega \approx \Delta_0$, where the effect of the coupling of
the TSS to the HO is most effective and therefore dephasing of the 
HO penetrates through to the TSS best. \\
The inset of Fig.\  \ref{fig.varom2} can be readily compared 
to the result of the recent 
work by Kleff {\em et al.\/}, see Fig.\ 3b of Ref.\ \cite{Kleff03}. 
There, the dephasing rate shows a particular almost discontinuous dependence 
on the ratio of $\Omega/\Delta_0$ around 1. 
In contrast to their findings, we do not observe such a type of 
behavior but instead a rather smooth dependence. The almost 
discontinuous feature is likely to be due to the fact that in Ref.\
\cite{Kleff03}, the zero temperature dynamics has been considered. 

%
%%%%%%%%%%%%%%%%%%%%%%%%%%%%%%%%%%%%%%%%%%%%%%%%%%%%%%%%%%%%%%%%%%%%%%%%%
%%%%%%%%%%%%%%%%%%%%%%%%%%%%%%%%%%%%%%%%%%%%%%%%%%%%%%%%%%%%%%%%%%%%%%%%%
%
\section{Varying the TSS-HO coupling strength $g$}
\begin{figure}[t]
\begin{center}
\epsfig{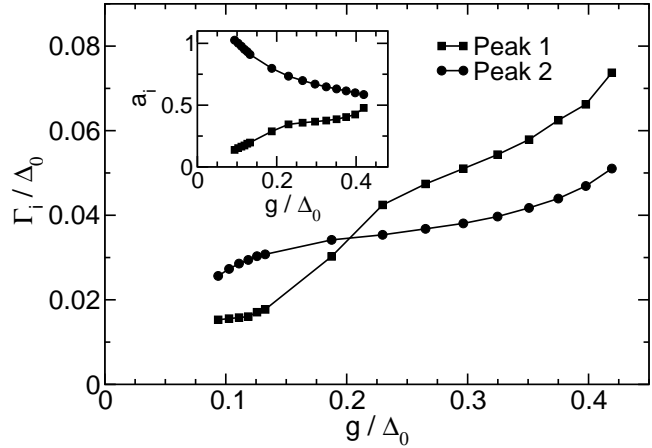}
\caption{Main: Dephasing rates in dependence of the coupling $g$ 
between the TSS and the HO for negative detuning $\Omega=0.75 \Delta_0$. 
For the remaining  parameters see Fig.\   \ref{fig.varom2}. Inset: 
The corresponding spectral weights $a_i$. One observes that 
peak 2 is dominating for this detuning. \vspace{8mm}
\label{fig.vara1}
}
\end{center}
\end{figure}

The dependence of the dephasing rates on
the coupling $g$ between the TSS and the HO is depicted in Fig.\  
\ref{fig.vara1}  for the case of 
negative detuning $\Omega< \Delta_0$. 
As expected, it increases monotonically with increasing 
$g$ and for this case, peak 2 is dominating, see the 
spectral weights $a_i$ in the inset of Fig.\  
\ref{fig.vara1}.   When the coupling $g$  between the TSS 
and the decoherent HO is increased, 
dephasing of the TSS is also larger. 
In general, the dependence of the dephasing on $g$ is observed 
to be rather weak, as long as $g< \Delta_0$, which is usually the case.  
%
%%%%%%%%%%%%%%%%%%%%%%%%%%%%%%%%%%%%%%%%%%%%%%%%%%%%%%%%%%%%%%%%%%%%%%%%%
%%%%%%%%%%%%%%%%%%%%%%%%%%%%%%%%%%%%%%%%%%%%%%%%%%%%%%%%%%%%%%%%%%%%%%%%%
%
\section{The role of temperature $T$}
\begin{figure}[t]
\begin{center}
\epsfig{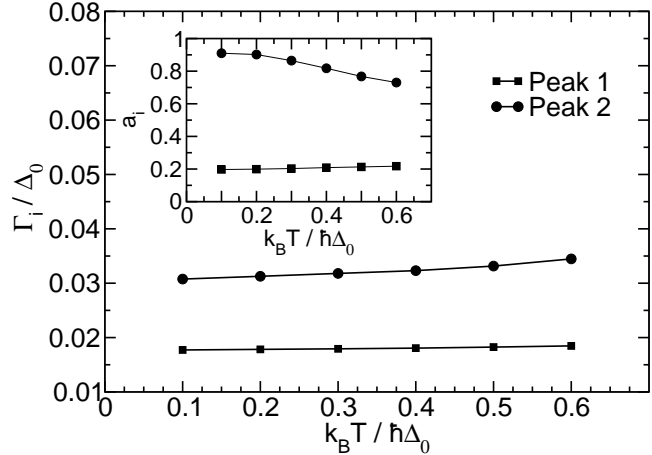}
\caption{Temperature dependence of the dephasing rates $\Gamma_1$ and
$\Gamma_2$ for the same parameters as in Fig.\ \ref{fig.varom2} (main) for 
$\Omega=0.75$. Inset: Corresponding spectral weights $a_i$. \vspace{5mm}
\label{fig.vartemp1}
}
\end{center}
\end{figure}

As in the pure Ohmic spin-boson system for weak damping, we also expect 
a rather weak dependence of the dephasing rate on temperature in the 
region of low $T$ for this more complicated system. Indeed, 
the results confirm this, see Fig.\  \ref{fig.vartemp1}. The dephasing
rates related to both peaks are almost constant with temperature $T$. 

%%%%%%%%%%%%%%%%%%%%%%%%%%%%%%%%%%%%%%%%%%%%%%%%%%%%%%%%%%%%%%%%%%%%%%%%%
%%%%%%%%%%%%%%%%%%%%%%%%%%%%%%%%%%%%%%%%%%%%%%%%%%%%%%%%%%%%%%%%%%%%%%%%%
%
\section{Dynamics in presence of a finite bias $\varepsilon\ne 0$}
\begin{figure}[t]
\begin{center}
\epsfig{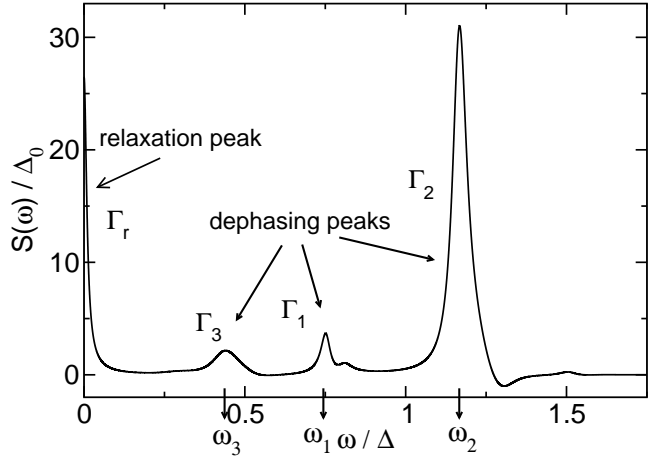}
\caption{Fourier transform $S_z(\omega)$ for the 
biased case $\varepsilon=0.5$. Compared to the unbiased case, two more 
resonances appear: the relaxation peak  of width $\Gamma_r$ at $\omega=0$
and a third dephasing peak of width $\Gamma_3$ at $\omega=\omega_3$. 
 Parameters are: $\Omega=0.75 \Delta_0, g=0.13 \Delta_0$, 
 $\kappa=0.014 (\rightarrow
\alpha = 0.003), k_{B}T=0.1 \hbar \Delta_0$. \vspace{9mm}
\label{fig.bias1}
}
\end{center}
\end{figure}

In this section, we consider the case of a biased TSS. Due to this bias, 
the spectrum shows two more resonances: (i) the relaxation peak at $\omega=0$, 
(ii) an additional dephasing peak. The frequency of the latter can be 
estimated by second order perturbation theory in $g$. It corresponds to the
energy difference between the states $|e,0\rangle$ and $|g,1\rangle$, which
is $\omega_3= \Delta_b-\Omega+\frac{2 g^2}{\Delta_b-\Omega}$. 
 A typical spectrum is shown in
Fig.~\ref{fig.bias1}. For this set of parameters, we obtain $\omega_3=0.46
\Delta_0$, which agrees well with the position of the numerically obtained 
resonance. The remaining two dephasing peaks at $\omega_{1,2}$ correspond 
to those of the unbiased case with the substitution $\Delta_0 \rightarrow 
\Delta_b$. 

\begin{figure}[t]
\begin{center}
\epsfig{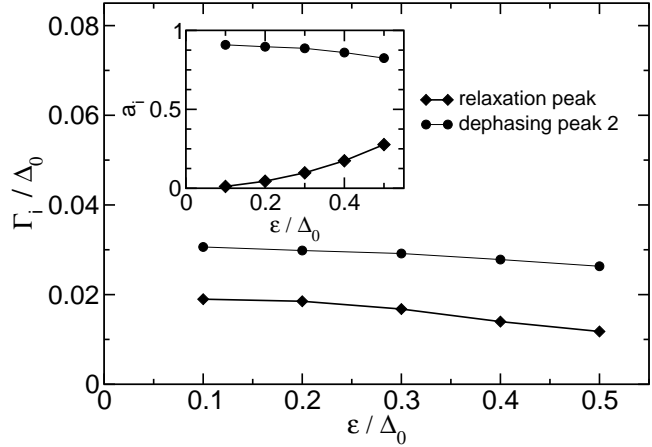}
\caption{Relaxation rate $\Gamma_r$ and dephasing rate $\Gamma_2$ 
for varying bias $\varepsilon$. Only the dominant dephasing peak 
is considered. The qualitative behavior of the other two 
rates is similar (not shown). 
 Parameters are: $\Omega=0.75 \Delta_0, g=0.13 \Delta_0$, 
 $\kappa=0.014 (\rightarrow
\alpha = 0.003), k_{B}T=0.1 \hbar \Delta_0$. Inset: 
Corresponding spectral weights $a_i$. \vspace{8mm}
\label{fig.bias2}
}
\end{center}
\end{figure}

The relaxation and the dephasing rates, $\Gamma_r$ and $\Gamma_{d_i}$,
respectively,  for increasing bias are shown in Fig.\  \ref{fig.bias2}. 
Note that 
we show only the dephasing rate of the dominant peak 2, i.e., of that one 
with the largest spectral weight. The qualitative behavior of the 
rates corresponding to the remaining two dephasing peaks is similar (not
shown). 
As one observes, the decay rates decrease with increasing bias for this 
case. This can be understood by considering the position of the resonance 
at $\Omega$ relative to the TSS  level spacing $\Delta_0$ for zero 
bias in the spectral density $J(\omega)$, Eq. (\ref{jeff}). For the  case 
shown in Fig.\ \ref{fig.bias2}, $\Omega < \Delta_0$. If we increase the bias
$\varepsilon$, the effective TSS  level spacing $\Delta_b$ increases. 
Since the spectral density is then decreasing (we move on the 
right flank of the resonance in Eq.\  (\ref{jeff}) towards larger 
$\omega$), dephasing is less effective
leading to decreasing decay rates. This reasoning  
can directly be verified  by considering the opposite situation 
$\Omega > \Delta_0$ (not shown). In this case, 
an increasing bias moves the TSS  level splitting towards  
increasing spectral weights, since we are moving on the left 
flank of the $\Omega$-resonance in Eq. (\ref{jeff}), causing 
an increase of the dephasing and relaxation.

%%%%%%%%%%%%%%%%%%%%%%%%%%%%%%%%%%%%%%%%%%%%%%%%%%%%%%%%%%%%%%%%%%%%%%%%%
%%%%%%%%%%%%%%%%%%%%%%%%%%%%%%%%%%%%%%%%%%%%%%%%%%%%%%%%%%%%%%%%%%%%%%%%%

\section{Approximate analytical approaches}

The standard analytical approach to the reduced dynamics of a
TSS weakly coupled to an environment consists in solving a  master
equation for the reduced density operator \cite{Cohen,Weiss99}. 
Path-integral methods \cite{Weiss99}, as well as the Bloch-Redfield 
formalism can for instance be used \cite{Argyres64}. 
The spin-boson system has been investigated in great detail for the 
case of weak-coupling to an Ohmic bath, having a continuous 
spectral density. The most general description dates back to 
Argyres and Kelly \cite{Argyres64} where also strong time-dependent 
driving has been included. Recently, the equivalence between the 
the path-integral approach and the Bloch-Redfield theory 
has been demonstrated in Ref.\  \cite{Hartmann00}.

In this section we shall apply a weak coupling approach to our problem, 
and compare with the exact numerical results presented in the previous 
sections.
In view of the possible application to more complex systems, it is in 
fact important to understand for this simple problem, in which regime 
the perturbative estimate differs from the exact solution. 
In addition, the weak-coupling approach is widely used to estimate the 
dephasing and relaxation rates in solid-state qubits 
\cite{vanderWaal00,Chiorescu03,Tian02}. 

In the second part of this section, we will show that a more
appropriate approach to the problem of decoherence due to a structured
environment consists in 
enlarging the Hilbert space of the quantum system 
in order to include the harmonic oscillator.

\subsection{Comparison with the spin-boson model for weak damping
\label{Sec.weakcoupl}}

\begin{figure}[t]
\begin{center}
\epsfig{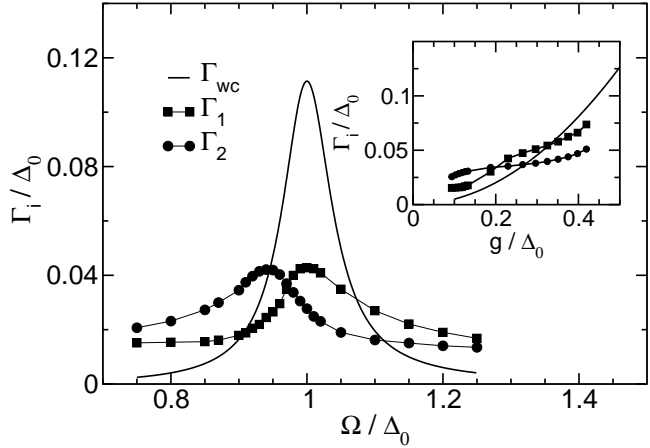}
\caption{Dephasing rates $\Gamma_1$ and  $\Gamma_2$ in comparison 
with the weak-coupling rate $\Gamma_{\rm d}$
for varying  $\Omega$.  
 Parameters are: $\varepsilon=0, g=0.07 \Delta_0$, 
$\kappa=0.014 (\rightarrow
\alpha = 0.0005$ for $\Omega=\Delta_0), k_{B}T=0.1 \hbar \Delta_0$. 
Inset: 
Corresponding results for varying $g$ for $\Omega=0.75 \Delta_0$. 
\label{fig.wc1}
}
\end{center}
\end{figure}

\begin{figure}[t]
\begin{center}
\epsfig{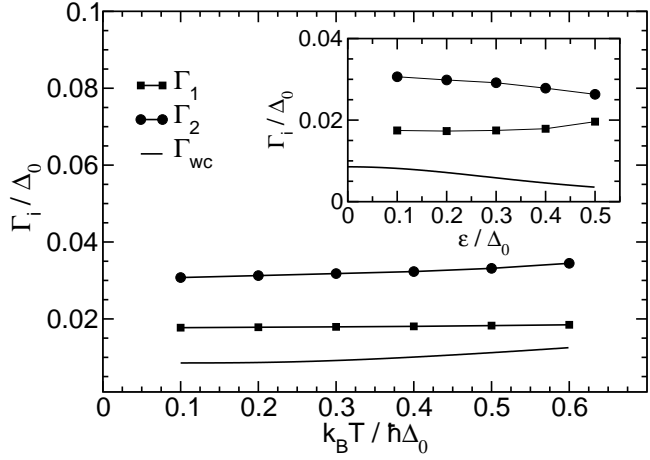}
\caption{Same as Fig.\  \ref{fig.wc1} for varying temperature $T$. 
 Parameters are: $\varepsilon=0, \Omega=0.75 \Delta_0, g=0.13 
 \Delta_0$, $\kappa=0.014 (\rightarrow
\alpha = 0.003)$.  Inset: 
Corresponding results for varying the bias $\varepsilon$ for 
$k_{B}T=0.1 \hbar \Delta_0$. 
\label{fig.wc2}
}
\end{center}
\end{figure}

For the spin-boson system the dephasing and relaxation rates can be evaluated
analytically to lowest order in the coupling to the bath
\cite{Weiss99,Grifoni97}. For the case of a spectral density showing Ohmic
behavior at low frequency, like Eq.(\ref{jeff}), they read
\begin{eqnarray}
\Gamma_d & = & \frac{\Gamma_r}{2} + \frac{\pi \varepsilon^2}{\Delta_b^2}
\frac{2 \alpha k_B T}{\hbar} \nonumber \\
\Gamma_r & = & \frac{\pi \Delta_{0}^2}{2 \Delta_b^2} J
(\Delta_b) \coth \frac{\hbar \Delta_b}{2 k_B T} \, ,
\label{wcrates}
\end{eqnarray}
where $\Delta_b = \sqrt{\Delta_{0}^2 + \varepsilon^2}$. 
The results for   the varying parameters $\Omega, g, T$ 
 are shown in
Figs.~\ref{fig.wc1} and \ref{fig.wc2} for the symmetric case. 
The dependence on the bias $\varepsilon$ is depicted 
in the inset of Fig.~\ref{fig.wc2}. 
We show only the results for the dephasing rates, 
the relaxation rate $\Gamma_r$ behaves in a qualitatively similar way. \\
We notice that the weak-coupling rates Eq.\  (\ref{wcrates}) give a 
good estimate of the dephasing rate only if $\Omega \gg \Delta_b$. 
This result can be understood considering the physics of the model
in the form given in Eq.\  (\ref{hamtlsosc}). 
If we regard the TSS as coupled to an environment represented by the 
damped HO, where $g$ is the coupling constant, the rates in 
Eq.\  (\ref{wcrates}) follow from the usual Bloch-Markov treatment.
This latter approach is valid when 
$g \, \tau_c \ll 1$, where $\tau_c$ represents 
the typical correlation times of the weakly damped oscillator \cite{Cohen}.
In the temperature regime considered $\tau_c$ may be identified with
the damping rate of the HO, $\gamma^{-1}$ \cite{Weiss99}.
Thus, the perturbative approach holds when $g \ll \gamma$. 
With the parameters of Fig.\ref{fig.wc1}, this condition is satisfied
for  ${\Omega \over\Delta_0} \gg 0.8$.

Of course the weak-coupling rates in Eq.\  (\ref{wcrates}) fail in
describing the dephasing rates if the TSS is close to or at 
resonance with the HO. 
For $\Omega \approx \Delta_0$, the weak-coupling rates in Eq.\ 
(\ref{wcrates}) overestimate the dephasing because there are coherent
exchange processes between the TSS and the HO mode which are not captured
in a weak-coupling approach. 
It is interesting to notice that for $\Omega < \Delta_0$ the damped HO 
behaves as an environment with long correlation times, and the weak coupling 
theory fails because of memory effects. 
The results in Eq.\  (\ref{wcrates}) underestimates the
dephasing rate in the regime $\Omega < \Delta_0$, 
a feature which has already been noticed for environments
responsible for  $1/f$ noise \cite{1overf}.

In fact, a weak-coupling approach is appropriate as long as the correlation
time of the bath is sufficiently short (as it is the case for an Ohmic bath
with small $\alpha$). However, the presence of the peak in the bath
spectral density induces bath correlations which decay only on a time scale
given by the width of the resonance.

\subsection{Three-level system as an approximative description\label{3ls}}
In order to obtain an analytical expression for the dephasing rates, we
restrict for simplicity to the simplest case of resonance 
$\Omega=\Delta_0$ between the TSS and the HO. Since the deviation from the
pure Ohmic case is most pronounced here, this is also the most 
interesting case for an analytical solution. 
We furthermore assume 
an unbiased TSS, $\varepsilon=0$. All the following considerations can be 
generalized to the off-resonant case and also with a biased TSS. The 
expressions, however, will become more involved. 
As it turns out, the TSS+HO can be restricted at low temperatures 
to a three-level-system (3LS) having performed second order perturbation theory in $g$. 
Assuming weak coupling to an Ohmic bath, we can derive the Redfield
equations which can be solved analytically. By this, we obtain closed 
expressions for the dephasing rates $\Gamma_i$ and the frequencies 
$\omega_i$.

\subsubsection{Redfield equations}

To start, we diagonalize the TSS+HO Hamiltonian which corresponds to the 
substitution $\sigma_x \rightarrow \sigma_z$ and $\sigma_z \rightarrow
-\sigma_x$ (the resulting transformed operators are denoted by the 
overline). 
Moreover, we apply the rotating wave approximation (RWA) which
is appropriate at resonance. Together with $\sigma_x = \sigma_+ + \sigma_-$, we
 obtain 
$\overline{H}^{RWA}_{TSS+HO}  =  -\frac{\hbar \Delta_0}{2} \sigma_z + \hbar \Omega B^{\dagger} B
-\hbar g (B\sigma_+ + B^{\dagger}\sigma_-)$. This is the Jaynes-Cumming
Hamiltonian which  can be  diagonalized exactly. 
The groundstate $|g,0\rangle$ is not shifted while the degeneracy between
the eigenstates of the uncoupled system is lifted. The higher lying eigenstates
appear in pairs where the energy of the $n$-th pair is $n\hbar\Omega$ above the 
ground state. Each pair is split by $\hbar\overline{\Delta}_n=2\hbar g \sqrt{n}$. 
As already used in Section \ref{sec.res}, we can restrict at low
temperatures and weak damping the system only to the ground-state 
$|g,0\rangle$ 
and the first pair of excited states, which are denoted as 
\begin{equation}
| -,0\rangle = \frac{1}{\sqrt2}[| e,0\rangle +| g,1\rangle ]
\;,\qquad
| +,0\rangle= \frac{1}{\sqrt2}[-| e,0\rangle +| g,1\rangle ] \, ,
\end{equation}
in the dressed states basis. The
corresponding energies of this 3LS are
$E_{-,0}=\hbar(\Omega -g)$ and
$E_{+,0}=\hbar(\Omega +g)$.  For our observable being the population difference,
this implies that $P_+(t)\equiv P(t)= \langle\sigma_z\rangle_t = 
tr_{TSS}(\rho(t) \sigma_z) = tr_{TSS}(\overline{\rho}(t) \sigma_x)$. In the
basis of the dressed states, we find
\begin{equation}
P(t)= \sqrt{2} {\rm Re }  \left( \langle g,0|\overline{\rho}(t)| -,0\rangle - 
\langle g,0|\overline{\rho}(t)| +,0\rangle \right) \, . 
\label{pt1}
\end{equation}
Hence, we need to solve the Redfield equations for the off-diagonal
elements $\overline{\rho}_{g0, -0}(t) 
\equiv \langle g,0|\overline{\rho}(t)| -,0\rangle$ and 
$\overline{\rho}_{g0, +0}(t) 
\equiv \langle g,0|\overline{\rho}(t)| +,0\rangle$. A standard 
derivation of the Redfield equations \cite{Weiss99} gives 
(to simplify the notation, we omit in the following the overline to 
denote the dressed states basis)
\begin{eqnarray}
\dot{\rho}_{g0, +0} & = & i (\Omega+g ) \rho_{g0, +0} - \frac{1}{2} 
\{C(\Omega+g)+C(\Omega-g)+C^*(-\Omega-g)\} \rho_{g0, +0} \nonumber \\ 
& & -\frac{1}{2} 
C^*(-\Omega+g) \rho_{g0, -0} 
+ \frac{1}{2} 
\{C(-\Omega-g)+C^*(\Omega+g)\} \rho_{+0, g0}\nonumber \\
& & +
\frac{1}{2} 
\{C(-\Omega+g)+C^*(\Omega+g)\} \rho_{-0, g0} \nonumber \\
\dot{\rho}_{g0, -0} & = & i (\Omega-g ) \rho_{g0, -0} - \frac{1}{2} 
\{C(\Omega+g)+C(\Omega-g)+C^*(-\Omega+g)\} \rho_{g0, -0} \nonumber \\ 
& & -\frac{1}{2} 
C^*(-\Omega-g) \rho_{g0, +0} 
+ \frac{1}{2} 
\{C(-\Omega-g)+C^*(\Omega-g)\} \rho_{+0, g0}\nonumber \\
& & +
\frac{1}{2} 
\{C(-\Omega+g)+C^*(\Omega-g)\} \rho_{-0, g0} \, .
\label{redfield1}
\end{eqnarray}
Here, $C(\omega)$ is the one-side Fourier transform of the bath
autocorrelation function $C(t)=\int_0^\infty d\omega' 
J_{\rm Ohm}(\omega')\left[ \coth \frac{\hbar \omega' \beta}{2} 
\cos \omega' t - i \sin \omega' t\right]$. 
It follows that 
\begin{eqnarray}
{\rm Re} C(\omega) & = & \frac{\pi}{2} J_{\rm Ohm}(\omega) [\coth \frac{\hbar \omega
\beta}{2} -1] \nonumber \\
{\rm Im} C(\omega) & = & \int_0^\infty d\omega'
\frac{J_{\rm Ohm}(\omega')}{\omega'^2-\omega^2} [\omega \coth \frac{\hbar \omega'
\beta}{2} -\omega'] \equiv \Lambda (\omega) - \Lambda' (\omega) \, .
\label{correlator}
\end{eqnarray}
In a next step, we use $C(\pm \Omega\pm g) \approx C(\pm \Omega)$ and
perform the secular approximation by neglecting terms in $\rho_{k,l}$ 
for which the Redfield tensor elements $|R_{nm,kl}| \ll
|\omega_{nm}-\omega_{kl}|$. In particular, this implies to neglect 
the terms with $\rho_{+0,g0}$ and $\rho_{-0,g0}$. 
We arrive at the final form of the Redfield equations %
\begin{eqnarray}
\dot{\rho}_{g0, +0} & = & i (\Omega+g ) \rho_{g0, +0} - \frac{\gamma_1}{2} 
 \rho_{g0, +0}  -\frac{\gamma_2}{2}  \rho_{g0, -0} \, ,\nonumber \\
\dot{\rho}_{g0, -0} & = & i (\Omega-g ) \rho_{g0, -0} - \frac{\gamma_1}{2} 
 \rho_{g0, -0}  -\frac{\gamma_2}{2}  \rho_{g0, +0} \, ,
 \label{redfield2}
\end{eqnarray}
where the rate constants are given by 
$\gamma_1=2C(\Omega)+C^*(-\Omega)$ and $\gamma_2=C^*(-\Omega)$.  
This system of coupled first-order differential equations can be solved by diagonalizing 
the coefficient matrix. The eigenvalues    
are given by 
\begin{equation}
\lambda_{\pm}  =  - \frac{\gamma_1}{2}  +i \Omega \pm 
\sqrt{\gamma_2^2 - 4 g^2}\, .
 \label{redfieldsol}
\end{equation}
The dephasing rates follow as $\Gamma_\pm = {\rm Re} \lambda_\pm$ and the 
oscillation frequencies are given by $\omega_\pm = {\rm Im} \lambda_\pm$. 
\\
The Redfield equations are expected to be applicable for weak damping,
i.e., $\kappa\ll 1$. Moreover, this 3LS approximation can only be applied
as long as the temperature is low enough, i.e., $k_{B}T \ll \hbar \Omega$. 
In addition, it is required that $\kappa \hbar \Omega \ll k_B T$. 
To verify our approximation, we compare in the following subsections the
results with the numerically obtained values of QUAPI.

\subsubsection{Dephasing rates}

Plugging in the Ohmic spectral density (\ref{johm}) in Eq.\ 
(\ref{correlator}), the dephasing rates $\Gamma_{\pm}$ follow as the real
parts of 
$\lambda_{\pm}$ in Eq.\  (\ref{redfieldsol}). We find   
\begin{equation}
\Gamma_{\pm} = \frac{\pi \kappa\Omega}{4} \left[ 3 \coth \frac{\hbar \Omega
\beta}{2} - 1\right] 
\pm \frac{1}{2} A \cos \frac{\varphi}{2} \, ,
\end{equation}
where 
\begin{equation}
\varphi= \arctan \frac{\pi \kappa\Omega \left[  \coth \frac{\hbar \Omega
\beta}{2} + 1\right]  I}{\frac{\pi^2}{4}\kappa^2 \Omega^2 \left[
\coth \frac{\hbar \Omega
\beta}{2} + 1\right]^2 - I^2 -4 g^2} \, ,
\end{equation} 
and
\begin{eqnarray}
A& =&  \left[ \left\{ \frac{\pi^2}{4} \kappa^2 \Omega^2 \left[  \coth \frac{\hbar \Omega
\beta}{2} + 1\right]^2 -   I^2 -4g^2 \right\}^2 \right.\nonumber \\ + 
&& \left.
\left\{ \pi \kappa\Omega \left[  \coth \frac{\hbar \Omega
\beta}{2} + 1\right]    I \right\}^2
\right]^{1/4} \, .
\end{eqnarray} 
Moreover, we have $I=\Lambda(\Omega)+\Lambda'(\Omega)$, where  
\begin{eqnarray}
\Lambda(\Omega) 
& = & \kappa\Omega \left\{ 
- {\rm Re } \psi \left( \frac{i \hbar \Omega}{2 \pi k_B T}  \right) 
+ \ln \left( \frac{\hbar \Omega}{2 \pi k_B T}  \right) \right.\nonumber \\
& & \left.
+ \frac{1}{2} 
\left[ e^{-\Omega/\omega_c} {\rm Ei} \left( \frac{\Omega}{\omega_c} \right) 
+  e^{\Omega/\omega_c} {\rm Ei} \left( -\frac{\Omega}{\omega_c} \right)  \right]
\right\} \,  ,
\end{eqnarray}
and
\begin{eqnarray}
\Lambda'(\Omega) 
& = & \kappa\left\{  \frac{1}{2}
\Omega 
\left[ e^{-\Omega/\omega_c} {\rm Ei} \left( \frac{\Omega}{\omega_c} \right) 
-  e^{\Omega/\omega_c} {\rm Ei} \left( -\frac{\Omega}{\omega_c} \right) -
\omega_c  \right]
\right\} \,  .
\end{eqnarray}
Here, $\psi(z)$ is the digamma function and ${\rm Ei}(z)$ denotes the exponential integral. 
Choosing the following dimensionless parameter set:
$\kappa=0.0014, T=0.01, \omega_c=10, g = 0.07  (\rightarrow \alpha=0.00005)$, we
obtain $\Gamma_+=0.0025$ and $\Gamma_-=0.002$, which have to be compared
with the QUAPI results $\Gamma_1=0.0030$ and $\Gamma_2=0.0027$. The 3LS
results deviate from the QUAPI results by $17\%
$ and $25\%
$, respectively. 
\subsubsection{Oscillation frequencies}

We find for the oscillation frequencies  
\begin{equation}
\omega_{\pm} = \Omega -\frac{3}{2} \Lambda(\Omega) + \frac{1}{2} \Lambda'(\Omega)
\pm \frac{1}{2} A \sin \frac{\varphi}{2} \, . 
\end{equation}
For the same parameter set as above, we obtain $\omega_+ =0.93$ and
$\omega_- =1.07$ from the 3LS and from QUAPI, we find $\omega_1=0.93$ and 
$\omega_2=1.07$. The difference between the two
rates is $\Delta \omega = 2 g$, as accurately predicted by the result
from the exact diagonalization of the Hamiltonian in RWA. 

%%%%%%%%%%%%%%%%%%%%%%%%%%%%%%%%%%%%%%%%%%%%%%%%%%%%%%%%%%%%%%%%%%%%%%%%%
%%%%%%%%%%%%%%%%%%%%%%%%%%%%%%%%%%%%%%%%%%%%%%%%%%%%%%%%%%%%%%%%%%%%%%%%%
%%%%%%%%%%%%%%%%%%%%%%%%%%%%%%%%%%%%%%%%%%%%%%%%%%%%%%%%%%%%%%%%%%%%%%%%%
%
\section{Application to experimentally realized superconducting 
flux qubits}
In this section, we apply our model to qubit devices 
which have been realized experimentally in the Delft group 
of J.\ Mooij. We refer to two devices, i.e., (i) 
the more recent flux 
qubit of Ref.\  \cite{Chiorescu03}, where the SQUID was directly attached 
to the qubit in order to increase the mutual coupling,  
 and (ii) the 
flux qubit of Ref.\  \cite{vanderWaal00} which was inductively 
coupled to a surrounding dc-SQUID being not directly in contact. \\
In order to extract the relevant parameters for our model, we
use the correspondence of the qubit-SQUID setup with an electronic
circuit \cite{Tian02}. This allows us to express the
parameters in our model in terms of the experimentally available parameters. 

\begin{figure}[t]
\begin{center}
\epsfig{figure=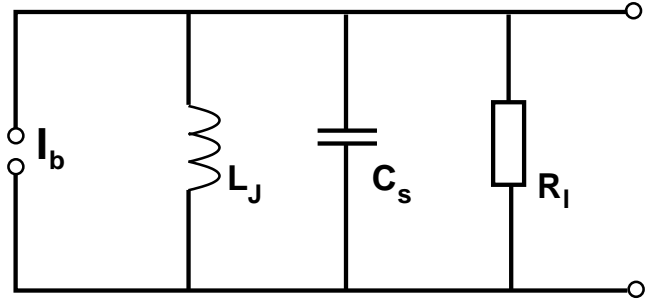,width=85mm,keepaspectratio=true}
\caption{Sketch of the circuit model of the dc-SQUID consisting of an ideal
inductance $L_J$ which is shunted by the capacitance $C_s$. The Ohmic
resistance is captured by $R_l$. For more details, see Ref.\  \cite{Tian02}. 
\label{fig.circ}
}
\end{center}
\end{figure}

In this approach, the damped dc-SQUID provides an electromagnetic
environment for the qubit. It can be characterized by the impedance of the
corresponding circuit sketched in Fig.\ \ref{fig.circ}.

The dc-SQUID far away from the switching point 
is modelled by an ideal inductance $L_J$. In the investigated
devices, a large shunt capacitance $C_s$ was designed in order to control
the environment. All Ohmic resitance of the circuit is modelled by $R_l$. 
Together with the flux quantum $\Phi_0$,
 the plasma frequency of the SQUID can be calculated as 
 $\omega_{p} = \frac{1}{\sqrt{L_J C_s}} = \sqrt{\frac{2 \pi \, 2 I_{c0} \cos f}{C_s \Phi_0}}$ .
Here, $I_{c0}$ is the critical current of a single Josephson junction 
(the SQUID possesses two junctions) and $f=\pi \Phi_{\rm ext} / \Phi_0$ is the 
scaled external flux, which is also denoted as the frustration of the
SQUID.  
By comparing our expression (\ref{jeff}) for the effective spectral density
with Eq.\ (7) in Ref.\ \cite{Tian02}, we find for our parameters:
\begin{equation}
\Omega  =  \omega_p \, , \mbox{\hspace{5ex}}
\kappa=  \frac{1}{C_s R_l} \frac{1}{2 \pi \omega_p} \, , \mbox{\hspace{5ex}}
\alpha  =  \frac{(2 e I_p I_b M_q)^2}{C_s^2 \hbar^3 R_l} \frac{1}{2
\omega_p^4}. 
\end{equation}
 For the coupling between 
qubit and SQUID, the persistent current $I_p$ of the qubit interacts via 
the mutual inductance $M_q$ with the bias current $I_b$ of the SQUID. 

An alternative way to obtain the Hamiltonian (\ref{hamtlsosc}) has been
worked out by H.\ Nakano and H.\ Takayanagi \cite{Nakano03} by starting
from a microscopic Hamiltonian. It carefully takes into account the 
phases for the  three Josephson junctions in the qubit ring and also the two 
junctions of the SQUID. By integrating out the corresponding hidden degree
of freedom and by approximating the SQUID potential well for low bias currents
by a harmonic oscillator, one arrives at a Hamiltonian equivalent to our
Eq.\  (\ref{hamtlsosc}). 
\subsection{Qubit 1}
For the Delft qubit of Ref.\  \cite{Chiorescu03}, 
the measured value of the level spacing is $\Delta_0 = 2 \pi  \, 3.4$
GHz. The persistent current of the qubit is $I_p= 330$ nA. The measurements
of the dephasing and relaxation times were done for the biased qubit 
with an effective level 
separation $\Delta_b=2 \pi \,  5.71$ GHz. 
The dc-SQUID is specified by the parameters $I_{c0}=2.2 \mu$A$, C_s =
10$ pF and  $f=\arccos
(\frac{1}{8})=1.45$.  The plasma frequency follows as $\omega_p=2 \pi  \,
2.0$ GHz. \\
In the Delft experiment, the inductive coupling between the qubit and the
SQUID was rather large, i.e.,  
$M_q = 9$ pH and the circuitry environment  is assumed to have an  Ohmic 
resistance $R_l  \approx 160 \Omega$.
The measurement time is therefore short. Hence, the experiment detected 
dephasing and relaxation of the qubit during operation 
 during which the qubit is in principle 
decoupled from the detector. However, due to the experimental design, the
decoupling is not perfect. A rather likely small asymmetry between the two 
Josephson junctions in the SQUID generates a small circulating current
which is compensated by a small bias current flowing through the SQUID. 
Assuming \cite{Chiorescuprivcom} that this bias current is about 5\% 
of the critical current $I_{c0}$ of a junction, this yields an effective bias 
current also during operation time of $I_b = 110$ nA. 
 The experiment has been performed at a
temperature of $T=25$ mK. \\
Plugging in the parameters and scaling them with respect to the 
pure level spacing $\Delta_0$, we arrive at the dimensionless parameters 
$\varepsilon=1.349, \Omega=0.59, \kappa=0.0078, g=0.008, T=0.15,
(\alpha=1.1 \times 10^{-5})$. We assume for the cut-off
frequency $\omega_c=10$. 
With similar QUAPI parameters as above ($\Delta t=0.06, M=12$ and
$K=2$), we find for the dephasing rate $\Gamma_d=0.0001 \Delta_0$, which 
implies a dephasing time of 
$\tau_d=470$ ns. It deviates by more than %one
%order of magnitude 
a factor of $20$ from the measured value of $\tau_{d, exp}=20$ ns. \\
In order to check the applicability of the weak coupling approach, 
we calculate the dephasing rate for the above parameter set. 
The weak-coupling dephasing time is calculated to be 
$\tau_{d, wc}=6.6$ $\mu$s. The weak coupling approach 
underestimates dephasing in this case by %a factor of 3. 
more than one order of magnitude. 
The determination of the relaxation rate for this specific parameter set 
is not possible with QUAPI since we would have to iterate the dynamics 
up to unrealistically long times. \\
The relatively large calculated dephasing and relaxation times present good
news for a possible future use of the qubit in a quantum computer. We have
assumed as noise source the Johnson-Nyquist noise from the electromagnetic 
environment. The reason for this rather large deviation from the
experimental data is that in the experiment an accidental 
additional resonance with an environmental mode occurred. This mode stems from 
a superconducting loop designed around the full qubit-plus-SQUID device. 
Its resonance frequency has been determined experimentally to be 
around $3$ GHz, which is close to the qubit level spacing at the 
degeneracy point. This kind of resonance will be avoided in the 
design of the next generation of the qubit which is currently 
ongoing \cite{Chiorescuprivcom}. An additional filtering of external electronic noise 
amplified  by the various measuring devices is also expected  to 
increase the measured dephasing time. 
\subsection{Qubit 2}
We have also calculated the dephasing rate due to Johnson-Nyquist noise for
the flux qubit reported in Ref.\  \cite{vanderWaal00}. 
The parameters for this qubit are $\Delta_0 = 2 \pi \,  0.66$ GHz and $I_p= 500$
nA.  We determine the dephasing rate at the degeneracy point, i.e.,
$\varepsilon=0$. 
The dc-SQUID is specified by the parameters $2I_{c0}=200 $nA$, C_s =
30$ pF and  $f=0.75$.  The plasma frequency follows as $\omega_p=2 \pi \, 0.61$ GHz. 
\\  
The device was designed in the way that the dc-SQUID enclosed the qubit. 
 The mutual inductance is $M_q = 8$ pH and the Ohmic 
resistance of the leads is estimated as 
$R_l=100 \Omega$. In this design, the dc-SQUID dominates the 
dephasing of the qubit since the mutual coupling is weak and the
measurement required to average over many switching events of the qubit. 
The bias current of the SQUID is taken to be close to the 
switching current, i.e.,  $I_b=120$ nA. The experiment was performed at 
$T=30 $mK. \\
The dimensionless parameters follow again from scaling with respect to 
the level spacing $\Delta_0$. We find $\Omega=0.93, \kappa=0.014, 
g=0.063$ ($\alpha=5 \times 10^{-4}$) and $T=0.95$. 
The dephasing rate is determined by QUAPI with the same parameters as
above for qubit 1. We find $\Gamma_d=0.05 \Delta_0$ yielding a dephasing time 
of $\tau_d = 5 $ns. Also in this case, we can compare this result with
the outcome of the standard weak-coupling approach which yields the 
value of $\tau_d = 5 $ ns. The measured value for the dephasing time 
is $\tau_{d, exp}=10$ ns.  
\section{Conclusions}
In summary, we have investigated the dynamics of the spin-boson problem for
the case when the frequency distribution of the bath shows a distinct
resonance at a characteristic frequency $\Omega$. We have mapped this model onto that  
of a two-state-system (TSS)  coupled to a single harmonic oscillator (HO) mode with
frequency $\Omega$, the latter being weakly coupled to an Ohmic bath. Since an
Ohmic bath induces rather fast decaying correlations at  finite
temperatures, the numerical method of the quasiadiabatic propagator 
 path-integral (QUAPI) can be applied by using the TSS+HO as the central
 quantum system exposed to Ohmic damping. This becomes treatable by QUAPI  
  as the later relies on the technique of cutting the
 correlations after a finite memory time. Having then studied the decay
 rates for dephasing and relaxation of a quantum superposition of
 eigenstates of the TSS, we found new features of the dynamics.
 This includes at least two inherent frequencies of the time evolution of
 the TSS, as well as pronounced increased dephasing and relaxation if the
 TSS is in resonance with the localized harmonic mode. 
  Our findings clearly demonstrate that the whole frequency spectrum of the peaked environment, 
  and not only its low frequency Ohmic part 
  may become relevant in determining the TSS relaxation dynamics, if 
  the frequencies of the TSS and the HO are comparable.   
In particular, we showed that, 
for a peaked environment close to resonance conditions, 
 the commonly used estimate of the decay and relaxation rates to lowest order in the 
 coupling between the TSS and the harmonic
 reservoir becomes inadequate.  The appropriate route is to 
 consider the TSS+HO as the quantum system, and then perform perturbation
theory in the coupling between such a system and the smooth Ohmic environment. 
 To this extent, 
  we established the Redfield equations for the unbiased, resonant 
 TSS-HO system. It
 turns out that this system can be reduced to a three-level system at 
 temperatures well below the characteristic frequency $\Omega$.
 Within the three-level system approximation, we have 
 obtained  analytic expressions for the relaxation and
 dephasing rates, which well agree with the numerical QUAPI rates. \\
Finally, we have applied the general model to
 experimentally realized superconducting flux qubits coupled to a dc-SQUID
 playing the role of the detector. We find considerably long dephasing
 times which represents an encouraging result for a possible use of these
 devices as qubits for future quantum information processors. 
\section{Acknowledgments}
It is a pleasure for the authors to dedicate this work to Prof.\ Uli Weiss 
on the occasion of his 60$^{\rm th}$ birthday. Over the years, we enjoyed
many fruitful discussions and collaborations with him, while he was teaching us many
interesting topics of physics. We also would like to thank for useful
discussions  I.\ Chiorescu, G. Falci, K.\ Harmans, Y.\ Nakamura,  H.\ Nakano,
F. Plastina and K.\ Semba. 
This work has been supported by the Deutsche Forschungsgemeinschaft 
DFG (M.T., Th820/1-1), by the Dutch Foundation for Fundamental Research on
Matter FOM (M.T., M.G.) and by the Telecommunication Advancement
Organization TAO of Japan (M.T.). 

\end{document}